\documentclass[12pt]{article}
\pdfoutput=1
\usepackage{graphicx}
\usepackage{epsfig,cite}
\usepackage{amssymb}
\usepackage{amsmath}
\usepackage{dsfont}
\usepackage{multirow}
\usepackage{color}
\usepackage{subfigure,amstext,alltt,setspace}
\usepackage{amsbsy}
\usepackage{comment}
\usepackage{fullpage}
\usepackage{array}
\usepackage{booktabs,multirow,tabularx}
\usepackage{hyperref}
\usepackage{slashed}
\usepackage{url}

\newcommand{\order}[1]{{\cal O}(#1)}

\allowdisplaybreaks[1]
\begin{document}

\begin{flushright}
TIF-UNIMI-2016-6\\
IPPP/16/61 \\
Cavendish-HEP-16/11 
\end{flushright}

\vspace*{.2cm}

\begin{center}
 {\Large \bf{Higgs production in bottom-quark fusion:\\
matching beyond leading order}}
\end{center}

\vspace*{.7cm}

\begin{center}
 Stefano Forte$^{1}$, Davide Napoletano$^2$ and Maria Ubiali$^{3}$
\vspace*{.2cm}

\noindent
{\it
  $^1$ Tif Lab, Dipartimento di Fisica, Universit\`a di Milano and\\ 
INFN, Sezione di Milano,
  Via Celoria 16, I-20133 Milano, Italy\\
  $^2$ Institute for Particle Physics Phenomenology,\\
  Durham University, Durham DH1 3LE, UK\\
  $^3$ Cavendish Laboratory, University of Cambridge,\\
  J.J. Thomson Avenue, CB3 0HE, Cambridge UK\\}

\vspace*{3cm}

{\bf Abstract}
\end{center}

\noindent
We compute the total cross-section for Higgs boson production in
bottom-quark fusion using the 
so-called FONLL method for the matching of a scheme in which the $b$-quark is
treated as a massless parton to that in which it is treated as a
massive final-state particle, and extend our previous results
to the case in which the next-to-next-to-leading-log five-flavor
scheme result is 
combined with the next-to-leading-order $\order{\alpha_s^3}$ four-flavor scheme computation. 

\pagebreak


Higgs production in bottom fusion, like 
any process involving bottom quarks at the matrix-element level, 
may be computed using two different factorization
schemes, often called   four- and five-flavour schemes for short. In the
four-flavour scheme (4FS), the bottom quark is treated as a massive
object, which is not endowed with a parton distribution (PDF), and it
decouples from QCD perturbative evolution, which is performed only
including the four lightest flavors and the gluon in the DGLAP
equations, and likewise it decouples from the running of $\alpha_s$ so
that $n_f=4$ in the computation of the QCD $\beta$ function.
In the five-flavour  scheme (5FS), instead, the bottom quark is treated on
the same footing as other quark flavors, there is a $b$ PDF, and
$n_f=5$ in both the DGLAP and renormalization-group equations.

For high enough scales, mass effects become negligible, collinear
logarithms related to $b$-quark radiation are large and must be
resummed, and the 5FS is always more accurate. On the
other hand, very close to the production threshold mass effects are
important while collinear logs are not large, and the 4FS is more accurate. 
In principle, a computation performed at high
enough perturbative order in the 4FS will reproduce
the 5FS result, while this is not the case for a
5FS computation, in which $b$-mass effects are never
included. 

In practice, however, for Higgs production in bottom fusion
the leading-order production diagram, which is $\order{\alpha_s^0}$ (parton
model) in the 5FS,  is $\order{\alpha_s^2}$ in the
4FS, so one must go to very high order indeed in the
5FS computation in order to reproduce 4FS results. In fact, in the 5FS, the cross section is known up to
NNLO~\cite{Harlander:2003ai} and in the  
4FS up to NLO~\cite{Dittmaier:2003ej,Dawson:2003kb}.
 Furthermore, the characteristic
scale for this process is necessarily rather higher than the $b$
production threshold, but perhaps rather lower than the Higgs mass
itself~\cite{Maltoni:2012pa,Lim:2016wjo}, and in a rather wide
range the 4FS and 5FS computations at the highest
available accuracy disagree by a sizeable amount, with the 5FS result
being significantly larger than the 4FS one, though they can be
brought to agree with a
very low scale choices, $\mu\lesssim m_H/4$.  
All this suggests that a
reliable computation of this process requires the use of a matched
scheme which combines the accuracy of the 4FS and 5FS results.

In the previous work~\cite{Forte:2015hba} we have implemented for this
process the so-called FONLL matched scheme, first proposed in
Ref.~\cite{Cacciari:1998it} for $b$ production and extended in
Ref.~\cite{Forte:2010ta} to deep-inelastic scattering: this method can
be used to combine 4FS and 5FS computations performed
at any given perturbative accuracy, retaining the accuracy of both,
i.e. in such a way that from the point of view of any of the two
computations that enter the combined results the terms which are added
are subleading.

In Ref.~\cite{Forte:2015hba}, this method was used to combine the 
next-to-next-to-leading order 5FS result with the
leading-order 4FS computation --- this particular
combination was called FONLL-A, corresponding to the lowest order at
which the 4FS and 5FS results have a non-vanishing
overlap.  The main result  was
that the FONLL-A result is generally quite close to the the 5FS
computation, and only acquires a small correction from the massive 
4FS terms, though this correction
has a scale dependence which is comparable to its absolutes size. This 
is unsurprising given that the 4FS calculation was only
included at leading order.

In order to pin down the precise size of the massive corrections it is
thus necessary to include the massive terms at least to
next-to-leading order. This is the purpose of the present paper:  we
include an
extra perturbative order to the 4FS result in comparison to
FONLL-A, thereby constructing the FONLL-B matched result (according to
the nomenclature introduced in Ref.~\cite{Forte:2015hba}). 
This amounts to combining both the
4FS and 5FS computations at the highest order available for both.

The basic idea of the FONLL  method is to expand out the
5FS computation, in which logarithms of $\mu^2_r/m_b^2$
and $\mu^2_f/m_b^2$ are resummed to all orders, in 
powers of the strong coupling $\alpha_s$, and replace them with their 
massive-scheme counterparts, up to the same order at which the
massive-scheme result is known. 
The combination then retains the logarithmic accuracy of the 5FS result one starts from (with the $b$ quark treated as massless), 
but now also has the fixed-order accuracy of the massive result, up to
the order which has been included. Henceforth, we consistently
use the notation N$^k$LL to refer to the resummed accuracy of the
5FS computation (i.e. by LL we mean a computation 
in which $\left(\alpha_s \ln\frac{m^2_b}{\mu^2}\right)$ is treated 
as order one), and by N$^k$LO to the fixed order at which the massive
4FS is performed. The FONLL-A scheme of Ref.~\cite{Forte:2015hba} is
thus NNLL+LO, while the FONLL-B combination considered here is NNLL+NLO.

The only
technical complication of the FONLL method is that the two
computations which are being combined are performed in different
renormalization and factorization schemes. This difficulty is overcome
by re-expressing $\alpha_s$ and PDFs in the 4FS
computation in terms of their 5FS counterparts, so that
one single $\alpha_s$ and set of PDFs is used everywhere. Once this is
done, the 4FS and 5FS computations can be simply
added, with overlapping terms subtracted in order to avoid double
counting: the result has the structure
\begin{equation}
\label{FONLL}
\sigma^{FONLL}=\sigma^{(4)}+\sigma^{(5)}-\sigma^{(4),(0)},
\end{equation}
in which $\sigma^{(4)}$ and $\sigma^{(5)}$ are respectively the 4FS and
5FS results, and $\sigma^{(4),(0)}$ is
their overlap.
The contributions to $\sigma^{(4),(0)}$ can be viewed and obtained 
either from
expansion of the 5FS computation up to finite order (thereby
extracting them from the 5FS result) or as the massless
limit of the massive computation (thereby
extracting them from the 4FS result) - with the caveat
that the 4FS result in the massive limit acquires
collinear singularities which in the 5FS are factorized
in the PDFs. 

In order to extend the results of Ref.~\cite{Forte:2015hba} to FONLL-B
we must thus first work out to one extra order in $\alpha_s$ the
expansion of 4FS expressions in terms of 5FS $\alpha_s$ and PDFs, and then, determine to one extra fixed
order in $\alpha_s$ the overlap term $\sigma^{(4),(0)}$ of Eq.~(\ref{FONLL}).

The first goal is achieved by writing
\begin{equation}
  \label{massive:1}
  \sigma^{(4)}=\tau_H\int_{\tau_H}^{1} \frac{dx}{x}\int_{\frac{\tau_H}{x}}^{1} \frac{dy}{y^2}\sum_{ij=q,g}f_{i}^{(5)}(x,Q^2)f_j^{(5)}\left(\frac{\tau_H}{x y},Q^2\right)B_{ij}\left(y,\alpha_s^{(5)}(Q^2),\frac{Q^2}{m_b^2}\right),
\end{equation}
where $f_{i}^{(5)}$ and $\alpha_s^{(5)}$ are 5FS PDFs
and $\alpha_s$, and the coefficients 
\begin{equation}
  \label{massive:exp}
  B_{ij}\left(y,\alpha_s^{(5)}(Q^2),\frac{Q^2}{m_b^2}\right)=\sum_{p=2}^N\left(\alpha_s(Q^2)\right)^pB_{ij}^{(p)}\left(y,\frac{Q^2}{m^2_b}\right)
\end{equation}
         are such that if
 $f_{i}^{(5)}$ and $\alpha_s^{(5)}$ are re-expressed in terms of 
$f_{i}^{(4)}$ and $\alpha_s^{(4)}$, then the expression of
$\sigma^{(4)}$ in the 4FS is recovered:
\begin{equation}
  \label{massive:2}
  \sigma^{(4)}=\tau_H\int_{\tau_H}^{1} \frac{dx}{x}\int_{\frac{\tau_H}{x}}^{1} \frac{dy}{y^2}\sum_{ij=q,g}f_{i}^{(4)}(x,Q^2)f_j^{(4)}\left(\frac{\tau_H}{x y},Q^2\right)\hat\sigma_{ij}\left(y,\alpha_s^{(4)}(Q^2),\frac{Q^2}{m_b^2}\right),
\end{equation}
with
\begin{equation}
  \label{massive:3}
\hat\sigma_{ij}\left(y,\alpha_s^{(4)}(Q^2),\frac{Q^2}{m_b^2}\right)=\
\sum_{p=2}^N\left(\alpha_s(Q^2)\right)^p\hat\sigma_{ij}^{(p)}\left(y,\frac{Q^2}{m^2_b}\right).
\end{equation}
Note that here and in the following discussion on the 4FS,
$\hat\sigma_{ij}^{(p)}$ refer to the partonic 
cross sections computed in the 4FS, as highlighted by their explicit
dependence on the ratio $Q^2/m^2_b$.

The expressions relating the 4FS and 5FS
PDFs up to $\order{\alpha^2_s}$ are given in
Ref.~\cite{Buza:1996wv}. They turn out to be trivial at $\order{\alpha_s}$,
so in our case it is only the redefinition of $\alpha_s$ (due to
changing $n_f$ by one unit) which has an effect. Explicitly, 
the non-vanishing $B_{ij}^{(k)}$ coefficients are 
at $\order{\alpha_s^2}$ 
\begin{align}
  B_{gg}^{(2)}\left(y,\frac{Q^2}{m^2_b}\right)      & = \hat{\sigma}_{gg}^{(2)}\left(y,\frac{Q^2}{m_b^2}\right) \\
  B_{q\bar{q}}^{(2)}\left(y,\frac{Q^2}{m^2_b}\right) & = \hat{\sigma}_{q\bar{q}}^{(2)}\left(y,\frac{Q^2}{m_b^2}\right)
\end{align}
while at $\order{\alpha_s^3}$ the redefinition of $\alpha_s$ contributes:
\begin{align}
  B_{gg}^{(3)}\left(y,\frac{Q^2}{m^2_b},\frac{\mu_R^2}{m_b^2},\frac{\mu_F^2}{m_b^2}\right) & = \hat{\sigma}_{gg}^{(3)}\left(y,\frac{Q^2}{m_b^2}\right) - \frac{2 T_R}{3\pi} \ln{\frac{\mu_R^2}{\mu_F^2}}\hat{\sigma}_{gg}^{(2)}\left(y,\frac{Q^2}{m_b^2}\right)\\
  B_{q\bar{q}}^{(3)}\left(y,\frac{Q^2}{m^2_b},\frac{\mu_R^2}{m_b^2},\frac{\mu_F^2}{m_b^2}\right) & = \hat{\sigma}_{q\bar{q}}^{(3)}\left(y,\frac{Q^2}{m_b^2}\right)- \frac{2 T_R}{3\pi} \ln{\frac{\mu_R^2}{m_b^2}}\hat{\sigma}_{q\bar{q}}^{(2)}\left(y,\frac{Q^2}{m_b^2}\right) \\
  B_{gq}^{(3)}\left(y,\frac{Q^2}{m^2_b}\right)      & = \hat{\sigma}_{gq}^{(3)}\left(y,\frac{Q^2}{m_b^2}\right) \\
  B_{qg}^{(3)}\left(y,\frac{Q^2}{m^2_b}\right)      & = \hat{\sigma}_{qg}^{(3)}\left(y,\frac{Q^2}{m_b^2}\right).
\end{align}

\begin{figure}[!htb]
\begin{center}
\includegraphics[width=0.9\textwidth,angle=0]{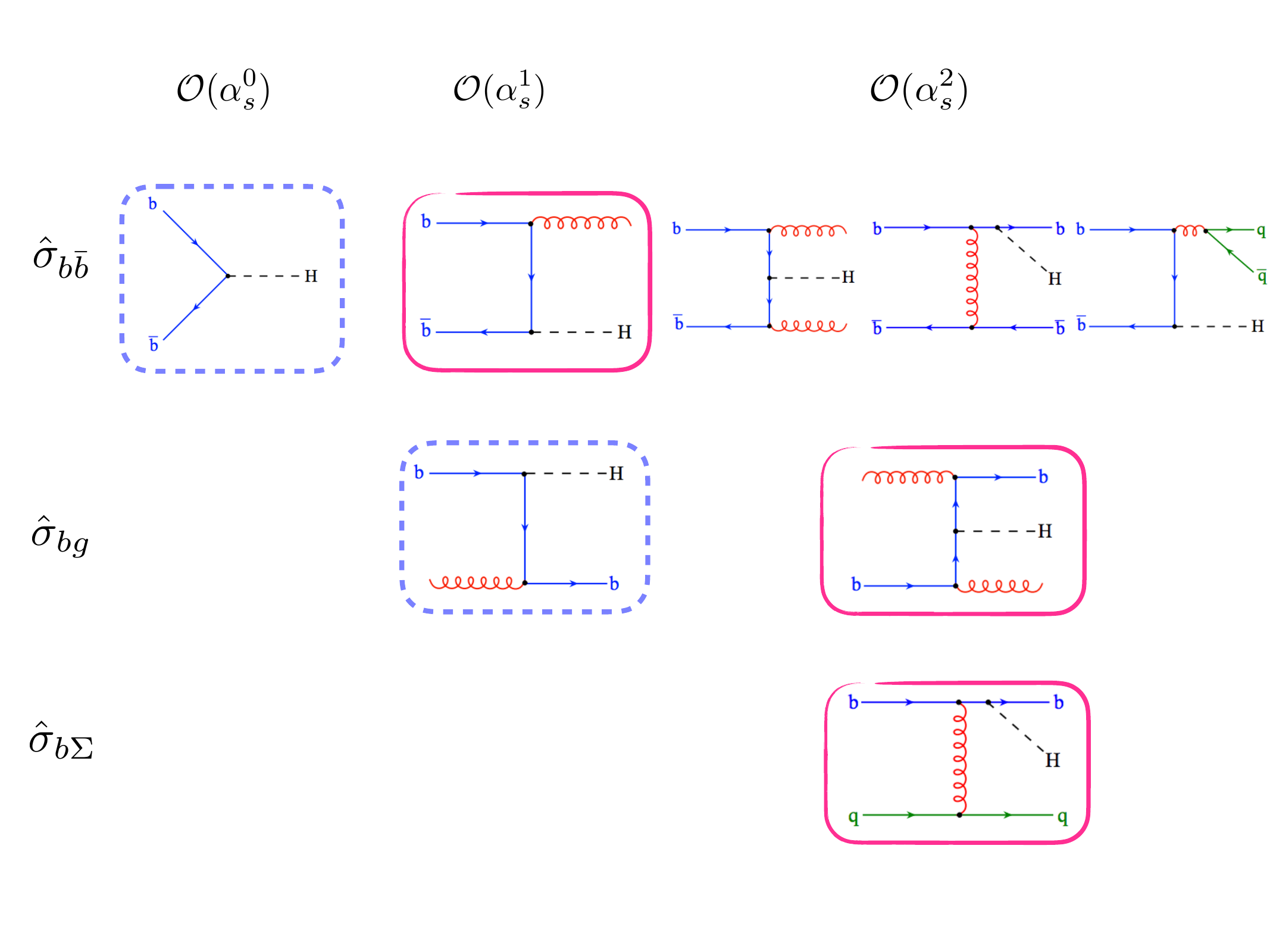} 
\caption{\label{fig:bbh} Representative examples of contributions to the 5FS
  computation which are subtracted and 
get replaced by massive 4FS
  contributions.  The diagrams circled with a dashed line become
  massive in FONLL-A, while those circled with a solid
  pink line are those that must be additionally subtracted in the
  FONLL-B scheme.}
\end{center}
\end{figure}

The second goal is achieved by
retaining all logarithms and 
constant terms in the 4FS NLO 
cross section and dropping all terms suppressed by powers of
$m_b/\mu$, namely by computing
\begin{equation}
  \label{eq:massless_lim}
  \sigma^{(4),(0)}\left(\alpha_s(Q^2),L\right)=\tau_H \int_{\tau_H}^{1} \frac{dx}{x}\int_{\frac{\tau_H}{x}}^{1} \frac{dy}{y^2}\sum_{ij=q,g}f_{i}(x,Q^2)f_j\left(\frac{\tau_H}{x y},Q^2\right)B_{ij}^{(0)}\left(y,L,\alpha_s(Q^2)\right),
\end{equation}
where 
\begin{equation}
L\equiv \ln Q^2/m_b^2
\end{equation}
and
\begin{equation}
  B_{ij}^{(0)}\left(y,L,\alpha_s(Q^2)\right) = \sum_{p=2}^N\left(\alpha_s(Q^2)\right)^pB_{ij}^{(0),(p)}\left(y,L\right),
\end{equation}
where the coefficients $B_{ij}^{(0),(p)}$ are defined as
\begin{equation}
  \lim_{m_b\rightarrow 0}\left[B_{ij}^{(p)}\left(y, \frac{Q^2}{m^2_b}\right)-B_{ij}^{(0),(p)}\left(y,\frac{Q^2}{m^2_b}\right)\right]=0.
\end{equation}

As already mentioned, all   $B_{ij}^{(0),(p)}$ terms in 
Eq.~(\ref{eq:massless_lim}) 
may be equivalently viewed as contributions to the 5FS computation, as
  schematically summarized in
Fig.~\ref{fig:bbh} (for real emission terms).
In
fact, 
because no simple closed-form expression of the massive coefficients
$B_{ij}^{(p)}$ is available,  it turns
out to be more convenient to extract the  $B_{ij}^{(0),(p)}$ from the
5FS result,  as it was done in Ref.~\cite{Forte:2015hba}. This is
simply done by expressing the 5FS $b$ PDF in terms of
the 4FS light quark and 
gluon PDFs up to $\order{\alpha_s}$ using the matching coefficients from 
Ref.~\cite{Buza:1996wv} (see also Appendix~C of
Ref.~\cite{Maltoni:2012pa}), and then re-expressing the result in
terms of the 5FS quark and gluon PDF, and 5FS $\alpha_s$. 

The result has the structure
\begin{align}
  \label{eq:btildesh}
  f_b^{(5)}(x,Q^2) = \alpha^{(5)}_s(Q^2) \int_{x}^1\frac{dz}{z} & \left\{ \mathcal{A}_{gb}^{(1)}\left(z,L\right)f^{(5)}_g\left(\frac{x}{z},Q^2\right)\vphantom{\frac{\alpha_s^{(5)}(Q^2)}{2\pi}} \right. \nonumber \\
  +& \left. \alpha_s(Q^2)\left[\mathcal{A}_{gb}^{(2)}\left(z,L\right)f^{(5)}_g\left(\frac{x}{z},Q^2\right) + \mathcal{A}_{\Sigma b}^{(2)}\left(z,L\right)f^{(5)}_{\Sigma}\left(\frac{x}{z},Q^2\right)\right]\right\}.
\end{align}
where $f^{(5)}_{b}$, $f^{(5)}_{\Sigma}$ and $f^{(5)}_{g}$ are
respectively the 5FS $b$ quark, singlet, and
gluon PDFs, and
\begin{align}
  \label{eq:btilde}
\mathcal{A}_{gb}^{(1)}&=a_{gb}^{(1,1)}(z)\, L \nonumber,\\
\mathcal{A}_{gb}^{(2)}&=a_{gb}^{(2,2)}(z)L^2 + a_{gb}^{(2,1)}(z) L + a_{gb}^{(2,0)}(z),\\
\mathcal{A}_{\Sigma b}^{(2)}&= a_{\Sigma b}^{(2,2)}(z)L^2 + a_{\Sigma b}^{(2,1)}(z) L + a_{\Sigma b}^{(2,0)}(z)\nonumber
\end{align}
Note that, as well known, to $\order{\alpha_s^2}$ the expression of the 5FS
$f^{(5)}_{b}$ in terms of the light quarks and gluon
receives constant (i.e. non-logarithmic) 
contributions $a_{g b}^{(2,0)}(z)$ and $a_{\Sigma b}^{(2,0)}(z)$, and
thus it is discontinuous at threshold $Q^2=m_b^2$ in the massless
scheme, as a consequence of it being continuous in the fully massive
calculation. The explicit expressions of the coefficients
Eq.~(\ref{eq:btilde}) are given in Appendix A for completeness.

We can now collect all contributions to $  \sigma^{(4),(0)}$.
The $\order{\alpha_s^2}$ terms, already given in
Ref.~\cite{Forte:2015hba}, are
\begin{align}
  B_{gg}^{(0)(2)} (y,L) & = y\int_y^1\frac{dz}{z}\left[2\mathcal{A}_{gb}^{(1)}\left(z,L\right)\mathcal{A}_{gb}^{(1)}\left(\frac{y}{z},L\right) + 4\mathcal{A}_{gb}^{(1)}\left(\frac{y}{z},L\right)\frac{1}{z}\hat{\sigma}_{gb}^{(1)}(z)\right] + \hat{\sigma}_{gg}^{(2)}(y), \\
  B_{q\bar{q}}^{(0)(2)} (y,L) &= \hat{\sigma}_{q\bar{q}}^{(2)}(y);
\end{align}
while the new contributions to  $\order{\alpha_s^3}$ are
\begin{align}\label{eq:subtrexp}
 B_{gg}^{(0)(3)} (y,L) & =
  y\int_y^1\frac{dz}{z}\left[4\mathcal{A}_{gb}^{(2)}\left(z,L\right)\mathcal{A}_{gb}^{(1)}\left(\frac{y}{z},L\right)\right.
    \nonumber\\
&    + 2\mathcal{A}_{gb}^{(2)}\left(\frac{y}{z},L\right)
\int_z^1\frac{dw}{w}   \mathcal{A}_{gb}^{(1)}\left(\frac{z}{w},L\right) \frac{1}{w} \hat{\sigma}_{b\bar{b}}^{(1)}(w)  \\
    &
\left.\phantom{asdfdy\int_y^1\frac{dz}{z}4\mathcal{A}_{gb}^{(2)}\left(z,L\right)}+
4\mathcal{A}_{gb}^{(2)}\left(\frac{y}{z},L\right)\frac{1}{z}\hat{\sigma}_{gb}^{(1)}(z) 
+
4\mathcal{A}_{gb}^{(1)}\left(\frac{y}{z},L\right)\frac{1}{z}\hat{\sigma}_{gb}^{(2)}(z)\right],\nonumber
  \\
  B_{gq}^{(0)(3)} (y,L)  & =
  y\int_y^1\frac{dz}{z}\left[2\mathcal{A}_{\Sigma
      b}^{(2)}\left(z,L\right)\mathcal{A}_{gb}^{(1)}\left(\frac{y}{z},L\right)
    + 2\mathcal{A}_{\Sigma
      b}^{(2)}\left(\frac{y}{z},L\right)\frac{1}{z}\hat{\sigma}_{gb}^{(1)}(z)
    \right.\nonumber  \\
    & \left.\phantom{asdfdy\int_y^1\frac{dz}{z}  4
      \mathcal{A}_{gb}^{(2)}\left(z,L\right)\mathcal{A}_{gb}^{(1)}\left(\frac{y}{z},L\right)\mathcal{A}_{gb}^{(1)}\left(\frac{y}{z},L\right)}+
    2\mathcal{A}_{gb}^{(1)}\left(\frac{y}{z},L\right)\frac{1}{z}\hat{\sigma}_{qb}^{(2)}(z)\right], 
\end{align}
which completes our result.
Note that in Eq.~(\ref{eq:subtrexp}) $\hat{\sigma}_{ij}^{(p)}(x)$ denotes
the partonic cross-section in the 5FS, as indicated by the fact that
it only depends on the momentum fraction and does not have any dependence 
on $m_b$ (unlike the
4FS partonic cross sections
$\hat{\sigma}_{ij}^{(p)}\left(x,\frac{Q^2}{m_b^2}\right)$ of
Eq.~(\ref{massive:3})). 

\begin{figure}
\begin{center}
\includegraphics[width=0.8\textwidth,angle=0]{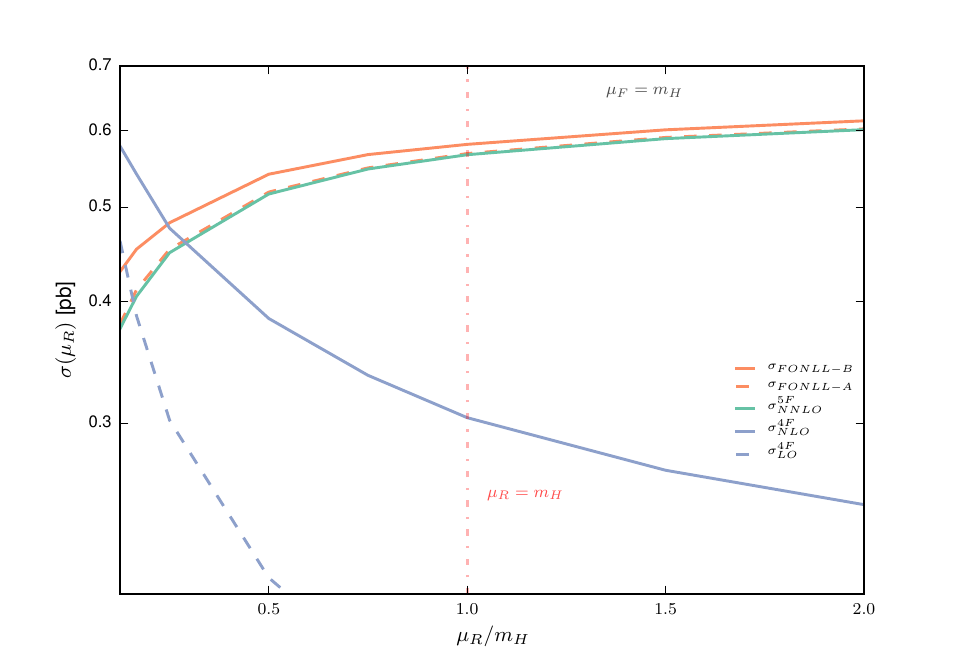} 
\includegraphics[width=0.8\textwidth,angle=0]{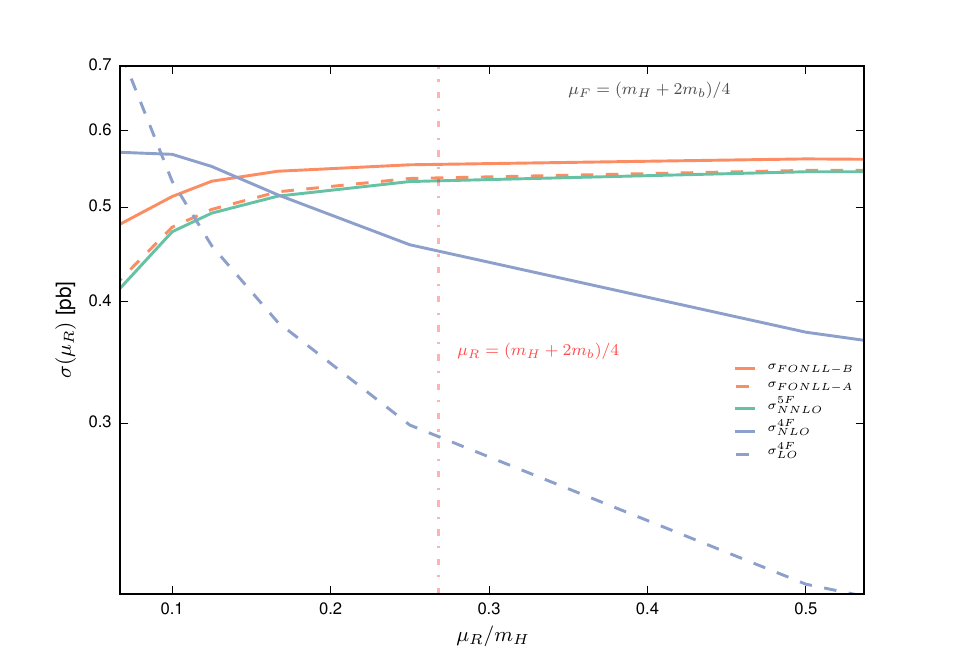}
\caption{\label{fig:muR_var} Comparison of the FONLL
  matched result and its 4FS and 5FS components,
  Eq.~(\ref{FONLL}). Results are shown as a function of the
  renormalization scale, with the factorization scale fixed at a high value 
$\mu_F=m_H$  (top) or a low value  $\mu_F=\frac{(m_H+2m_b)}{4}$ (bottom).}
\end{center}
\end{figure}

We have implemented our final FONLL-B expression by combining,
according to Eq.~(\ref{FONLL})
4FS predictions up to NLO 
obtained using MG5\_aMC@NLO~\cite{Alwall:2014hca,Wiesemann:2014ioa}, 5FS computations up
to NNLL obtained using  the {\tt bbh@nnlo}
code~\cite{Harlander:2003ai}, and our own implementation of the
subtraction term Eq.~(\ref{eq:massless_lim}).

\begin{figure}
\begin{center}
\includegraphics[width=0.8\textwidth,angle=0]{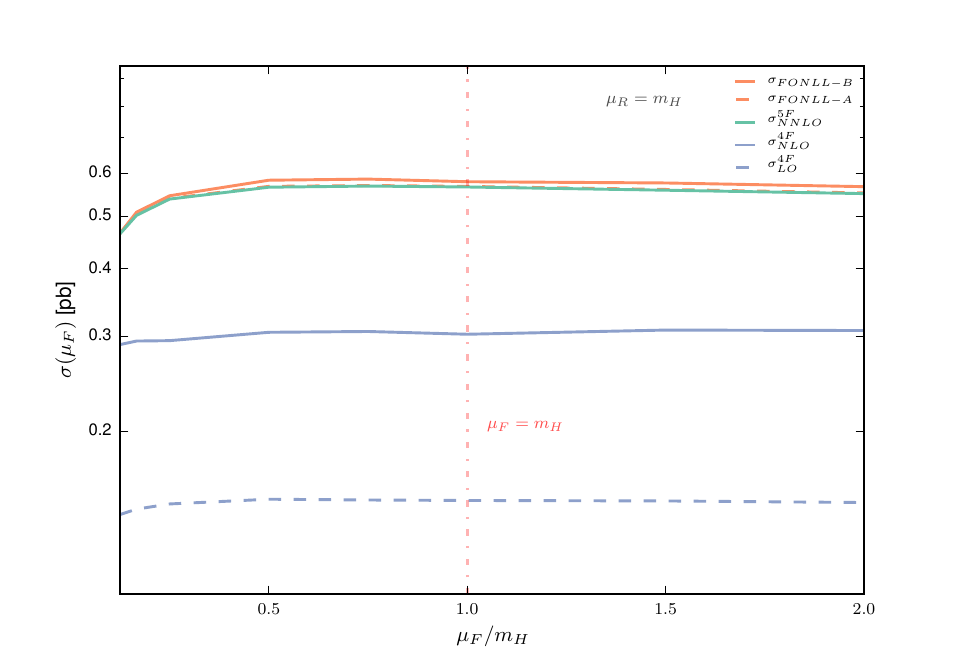} 
\includegraphics[width=0.8\textwidth,angle=0]{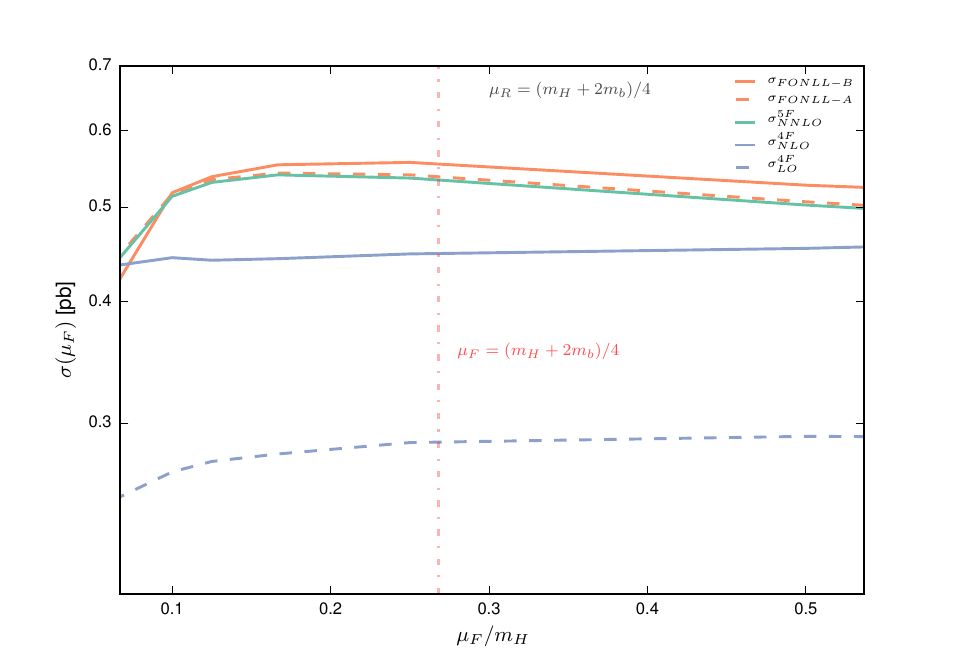}
\caption{\label{fig:muF_var} Same as Fig.~\ref{fig:muR_var}, but now
  with the factorization scale varied with the renormalization scale
  kept fixed at a high value 
$\mu_R=m_H$  (top) or a low value  $\mu_R=\frac{(m_H+2m_b)}{4}$ (bottom) .}
\end{center}
\end{figure}

\begin{figure}
\begin{center}
\includegraphics[width=0.8\textwidth,angle=0]{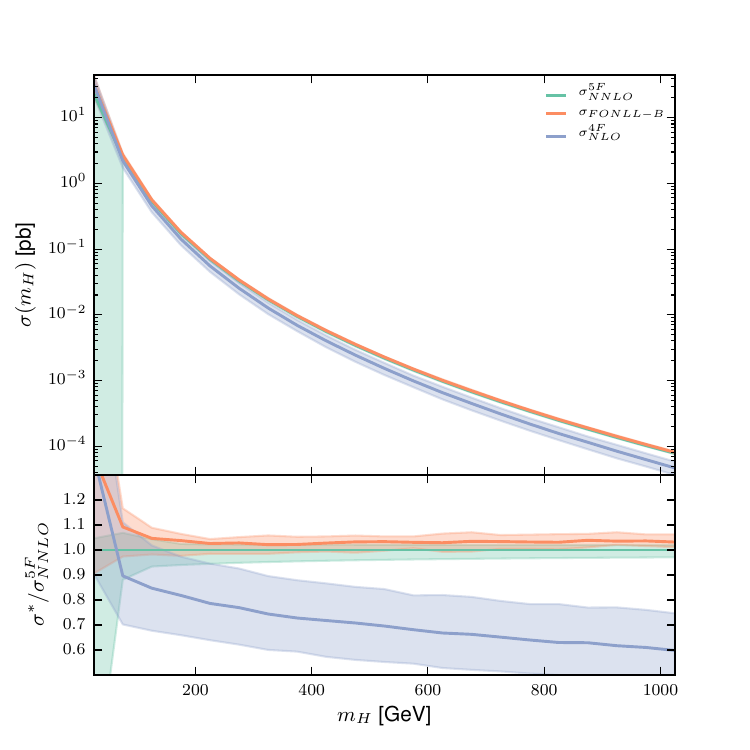} 
\caption{\label{fig:massdep} The cross-section using the most accurate
  results in the 4FS (NLO), 5FS (NNLO) and matched (FONLL-B) schemes,
  as a function of the Higgs mass, with
  $\mu_r=\mu_F=\frac{(m_H+2m_b)}{4}$. The bottom panel shows the
  result as a ratio to the 5FS computation. The uncertainty band is
  obtained by standard seven-point scale variation (see text).}
\end{center}
\end{figure}

In Figs.~\ref{fig:muR_var}-\ref{fig:massdep} we  compare the 4FS, 5FS
and matched FONLL results. Specifically, in
Figs.~\ref{fig:muR_var}-\ref{fig:muF_var} we show for the physical
Higgs mass value $m_H=125$~GeV, varying the renormalization and factorization
scale
 both the LO and NLO
4FS predictions, and the FONLL-A and FONLL-B matched results in which
they are respectively combined with the NNLL 5FS result, also
shown. In  Fig.~\ref{fig:massdep} we show  the most accurate
results obtained in the 4FS (NLO), 5FS (NNLO) and matched (FONLL-B)
schemes, as a function of the Higgs mass, with
$\mu_R=\mu_F=\frac{m_H+4m_B}{4}$, and the uncertainty band obtained by
taking the  envelope of the
variations of  the renormalization and factorization scales by a factor two
about the central value with the two outer points $\mu_R=4\mu_F$ and
$\mu_F=4\mu_r$ omitted. Note that for the lowest (unphysical) Higgs
mass values this uncertainty blows up because the lower edge of the
scale variation range extends in the nonperturbative region.

The 4FS results shown  are  those which enter the FONLL
combination, namely, the form Eq.~(\ref{massive:1}) of the 4FS result
is used, in which this is expressed in terms of 5FS PDFs and $\alpha_s$.
 All results are computed using a PDF set presented and
discussed in Ref.~\cite{Bonvini:2016fgf}. This PDF set is based on the 
PDF4LHC15 combined
sets~\cite{Butterworth:2015oua,Ball:2014uwa,Harland-Lang:2014zoa,Dulat:2015mca,Carrazza:2015aoa,Gao:2013bia,Watt:2012tq},
with which it is taken to coincide below the $b$ mass, but from which
it is then evolved up in the $5FS$ from $Q=m_b$, with the results
below and above threshold matched exactly as in
Eq.~(\ref{eq:btildesh}). This is not quite the 
same as the original PDF4LHC15 combination, which is obtained by
combining sets which adopt different values of $m_b$, and also
incorporate
subleading
differences in the way the 4FS and 5FS are matched at threshold: it
thus has the advantage of being fully consistent. We use pole-mass
expressions and take a $b$ pole-mass value
$m_b=4.58$~GeV; the strong coupling is $\alpha_s(m_Z)=0.118$. 

From Fig.~\ref{fig:muR_var} we see that the strong renormalization scale
dependence of the LO 4FS result is reduced at NLO, and also,
that at NLO the big gap between the 4FS and 5FS results gets compensated for
by the inclusion of higher order terms in the 4FS. 
This, together with the fact that the 5FS shows very little
scale dependence, and that differences are significantly smaller for smaller values of
$\mu_F$, strongly suggests that the bulk of the difference
between the 4FS and the 5FS is due to large logs of
$\mu_F^2/m_b^2$ which are resummed into the PDF in the latter case. 
This is in agreement with the conclusion of Ref.~\cite{Lim:2016wjo},
in which it was shown that resummation increases
the cross section in most cases by up to 30\% at the LHC, leading to a better precision. 
On the other hand, the 4FS predictions at NLO also displays a consistent
perturbative behaviour only when evaluated at a suitably low scale.

The massive corrections which the 4FS result contains turn out to be 
 much smaller, though not entirely negligible. Indeed, whereas the FONLL-A result
essentially coincides with the 5FS, the FONLL-B, which only differs
from it
because of the inclusion of massive terms at one extra perturbative
order, departs somewhat from it\footnote{In Ref.~\cite{Forte:2015hba} the FONLL-A result
, while also close to the 5FS result, did not coincide exactly with it
  for generic scales, because their respective scale dependences,
  though slight, had different shapes. This difference in
shape was due to the fact that, unlike here, a fully consistent PDF
set was not used: rather, the PDFs were taken from a public set, with
a value of $m_b$ which differed from that used in the computation of
the matrix element, thereby leading to a mismatch in the scale dependence.}.
The factorization scheme dependence
shown in Fig.~\ref{fig:muF_var} is very mild in all schemes when
$\mu_R$ is high, but for low $\mu_R$, where the perturbative expansion
of the 4FS result is more reliable, both the 5FS and the FONLL-A results
show a contained scale dependence, comparable in size to the mass effects,
which is reduced in the FONLL-B result.

These results  suggest that the main  difference
between the FONLL-A and the FONLL-B schemes is the inclusion of 
a higher order contribution from the 4FS computation which reduces the 
the scale dependence of the FONLL-A result; because the latter is
essentially the same as that of the 5FS computation this contribution
is likely to be dominated by  a constant, i.e., mass-independent
term. 
This conclusion is
supported by Fig.~\ref{fig:massdep}, in which results are shown as a
function of the Higgs mass: the difference between the FONLL-B and 5FS
results decreases slightly as the mass grows until $m_H\sim200$~GeV,
but then it remains constant up to the highest values of the Higgs
mass.

We conclude that the FONLL-B result is the most reliable, and a low
choice of renormalization and factorization scheme seems to lead to a
more reliable perturbative expansion, but all in all mass corrections
are very moderate, so the usage of the 5FS result at all scales would be
adequate in most cases. This rather disfavours phenomenological
combinations such as the so-called Santander
matching~\cite{Harlander:2011aa} in which the 4FS and 5FS results are
combined through an interpolation that gives each of them comparable
weight. The difference between the FONLL-B and 5FS is almost entirely
due to a constant $O(\alpha_s^3)$ mass-independent contribution which
appears in the 4FS at NLO but would only enter the 5FS at
N$^3$LO; the FONLL-B computation, which includes it, is accordingly
more accurate, even for very high vales of the Higgs mass.

Matched results for this process were recently obtained in
Refs.~\cite{Bonvini:2015pxa,Bonvini:2016fgf} using an
effective field theory approach,
and a somewhat different counting of perturbative orders. A benchmarking
of our results with those of these references has been performed in the
context of the Higgs cross section working group, and it will be
presented there~\cite{Anastasiou:xxx}. In that  benchmarking 
the matched calculations are found to agree to within better than
$5\%$ when results at the same
perturbative orders  are included, with the residual difference due
to a somewhat different choice of factorization and renormalization
scales in the two computations which are being compared.

In summary, we have presented a matched computation of Higgs
production in association with bottom quarks including known results
to the highest available accuracy, namely, NLO in a four-flavor scheme
in which $b$ quark mass effects are fully accounted for, and  NNLL in
a five-flavor scheme in which the $b$ quark is treated as a massless
parton with collinear logs resummed to all orders.
We find that mass corrections are very small while collinear logs are
substantial, so that in practice the fully matched result is very
close to the 5FS one. The fully matched result receives a small
correction from mass effects and it is very stable upon renormalization
and factorization scheme variation, suggesting that it is adequate for
precision phenomenology at the LHC.

\bigskip

A public implementation of our  NNLL+NLO FONLL-B matched computation
will be made available from:
\begin{center}
\url{http://bbhfonll.hepforge.org/}
\end{center}

\section*{Acknowledgements}
We thank Fabio Maltoni, Giovanni Ridolfi and Paolo Nason for illuminating discussions. 
We thank Marius Wiesemann for his help in comparing our results 
to those of Ref.~\cite{Wiesemann:2014ioa}, and Marco Bonvini, Andrew Papanastasiou
and  Frank Tackmann for
discussions on the their approach and on the PDFs of
Ref.~\cite{Bonvini:2016fgf}.
SF and DN are  supported by the European Commission through the
HiggsTools Initial Training Network  PITN-GA2012-316704. We thank
Claude Duhr, Falko Dulat, Valentin Hirschi and Berhard Mistlberger for
pointing out several typos in the published version of this paper.
\begin{appendix}
\section{Appendix A}
\numberwithin{equation}{section}
\setcounter{equation}{0}
We give for completeness the expressions of the coefficients
Eq.~(\ref{eq:btilde}). These were computed in Ref.~\cite{Buza:1996wv}.
There are a few differences compared to what is presented there.
Firstly we separate contributions from $b$ and $\bar{b}$. Secondly
our expansion is done in powers of $\alpha_s$ rather than in powers
of $\frac{\alpha_s}{4 \pi}$. Lastly we have re-expressed the gluon
and singlet PDFs in the 4FS in terms of those computed in the 5FS.
\begin{align}
\mathcal{A}^{(2)}_{\Sigma b}(z,L) & =
\frac{1}{32 \pi^2 }C_FT_f\Biggl\{
\Biggl[-8(1+z)\ln z-\frac{16}{3z}-4
+ 4 z +\frac{16}{3}z^2\Biggr] L^2
\nonumber \\ &  
-\Biggl[8(1+z)\ln^2z-\Biggl(8+40z+\frac{64}{3}z^2\Biggr)\ln z
-\frac{160}{9z}
+16-48z+\frac{448}{9}z^2\Biggr] L
\nonumber \\ &  
+ (1+z)\Biggl[32{\rm S}_{1,2}(1-z)+16\ln z{\rm Li}_2(1-z)
-16\zeta(2)\ln z
-\frac{4}{3}\ln^3z\Biggr]
\nonumber \\ &  
+\Biggl(\frac{32}{3z}+8-8z-\frac{32}{3}z^2\Biggr) {\rm Li}_2(1-z)
+ \Biggl( -\frac{32}{3 z}-8+8z+\frac{32}{3} z^2\Biggr)\zeta(2)
\nonumber \\ &  
+\Biggl(2+10z+\frac{16}{3}z^2\Biggr)  \ln^2z
-\Biggl(\frac{56}{3}+\frac{88}{3}z
+\frac{448}{9}z^2\Biggr)\ln z
-\frac{448}{27z} - \frac{4}{3}
-\frac{124}{3}z+\frac{1600}{27}z^2 \Biggr\}  \,,
\end{align}
\begin{align}
\mathcal{A}_{gb}^{(1)} (z,L) & = \frac{T_f}{2\pi} \Biggl[ ( z^2 + (1 - z)^2) L\Biggr] \, ,
\end{align}
and

\begin{align}
\!\!\!\!\!\!\!\!\!\!\!\!\!\!\!\!\!\!\mathcal{A}_{bg}^{(2)}(z,L) & =
\frac{1}{32 \pi^2 }\Biggl\{\Biggl\{C_FT_f[ (8 -16 z+16 z^2)\ln(1-z)
-(4 -8 z+ 16 z^2)\ln z -(2 - 8 z)]
\nonumber \\ &  
+C_AT_f\Biggl[-(8 - 16 z + 16 z^2)\ln(1-z)
-(8 + 32 z)\ln z
 -\frac{16}{3z} -4  - 32 z+\frac{124}{3}z^2\Biggr]\Biggr\} L^2
\nonumber \\ &  
-\Biggl\{C_FT_f \Biggl[( 8 - 16 z + 16z^2)[2\ln z\ln(1-z)
-\ln^2(1-z)+2\zeta(2)]
\nonumber \\ &  
-(4 - 8 z +16 z^2)\ln^2z-32z(1-z)\ln(1-z)
-(12 - 16 z + 32 z^2)\ln z  - 56+116z -80z^2 \Biggr]
\nonumber \\ &  
+ C_AT_f\Biggl[(16 +32 z +32 z^2)[{\rm Li}_2(-z) + \ln z\ln(1+z) ]
+(8 - 16 z + 16 z^2)\ln^2(1-z)
\nonumber \\ &  
+(8 + 16 z)\ln^2z
+32z\zeta(2)+32z(1-z)\ln(1-z)
-\Biggl(8+64z+\frac{352}{3}z^2\Biggr)\ln z
\nonumber \\ &  
-\frac{160}{9z}+16-200z+\frac{1744}{9}z^2 \Biggr]\Biggl\}L
\nonumber \\ &  
+ C_FT_f \Biggl\{(1-2z+2z^2)  [8\zeta(3)
+\frac{4}{3}\ln^3(1-z)
-8\ln(1-z){\rm Li}_2(1-z)
+8\zeta(2)\ln z
\nonumber \\ &  
-4\ln z\ln^2(1-z)
+\frac{2}{3}\ln^3z
-8\ln z{\rm Li}_2(1-z)
+8{\rm Li}_3(1-z)
-24{\rm S}_{1,2}(1-z)]
\nonumber \\ &  
+z^2\Biggl[-16\zeta(2)\ln z+\frac{4}{3}\ln^3z
+16\ln z{\rm Li}_2(1-z)+32{\rm S}_{1,2}(1-z)\Biggr]
\nonumber \\ &  
-(4+96z-64z^2){\rm Li}_2(1-z)
-(4-48z+40z^2)\zeta(2)
\nonumber \\ &  
-(8+48z-24z^2)\ln z\ln(1-z)
+(4+8z-12z^2)\ln^2(1-z)
\nonumber \\ &  
-(1+12z-20z^2)\ln^2z-(52z-48z^2)\ln(1-z)
\nonumber \\ &  
-(16+18z+48z^2)\ln z
+26-82z+80z^2\Biggr\}
\nonumber \\ &  
+C_AT_f\Biggl\{(1-2z+2z^2) [
-\frac{4}{3} \ln^3(1-z)
\nonumber \\ &  
+8\ln(1-z){\rm Li}_2(1-z)-8{\rm Li}_3(1-z)]
+(1+2z+2z^2)
\nonumber \\ &  
\times [-8\zeta(2)\ln(1+z)
-16\ln(1+z){\rm Li}_2(-z)
-8\ln z\ln^2(1+z)
\nonumber \\ &  
+4\ln^2z\ln(1+z) + 8\ln z{\rm Li}_2(-z)-8{\rm Li}_3(-z)
-16{\rm S}_{1,2}(-z)]
\nonumber \\ &  
+(16+64z)[2{\rm S}_{1,2}(1-z)
+\ln z{\rm Li}_2(1-z)]
-\Biggl(\frac{4}{3} +  \frac{8}{3} z\Biggr)\ln^3z
\nonumber \\ &  
+(8-32z+16z^2)\zeta(3)-(16+64z)\zeta(2)\ln z+(16+16z^2)
\nonumber \\ &  
\times [ {\rm Li}_2(-z) + \ln z\ln(1+z)  ]
+\Biggl(\frac{32}{3z}+12+64z-\frac{272}{3}z^2\Biggr)
{\rm Li}_2(1-z)
\nonumber \\ &  
-\Biggl( 12 + 48 z - \frac{260}{3} z^2+\frac{32}{3 z}\Biggr)\zeta(2)
-4z^2\ln z\ln(1-z)
\nonumber \\ &  
-(2+8z-10z^2)\ln^2(1-z)+\Biggl(2+8z+\frac{46}{3}z^2\Biggr)\ln^2z
\nonumber \\ &  
+(4+16z-16z^2)\ln(1-z)
-\Biggl(\frac{56}{3}+\frac{172}{3}z+\frac{1600}{9}z^2\Biggr)\ln z
\nonumber \\ & 
-\frac{448}{27z}-\frac{4}{3}-\frac{628}{3}z
+\frac{6352}{27}z^2\Biggr\}\Biggr\} \, .
\end{align}

\end{appendix}
\renewcommand{\em}{}
\bibliographystyle{UTPstyle}
\bibliography{bbH_FONLL.bbl}

\providecommand{\href}[2]{#2}\begingroup\raggedright\begin{thebibliography}{10}

\bibitem{Harlander:2003ai}
R.~V. Harlander and W.~B. Kilgore, {\it {Higgs boson production in bottom quark
  fusion at next-to-next-to leading order}},  {\em Phys. Rev.} {\bf D68} (2003)
  013001, [\href{http://xxx.lanl.gov/abs/hep-ph/0304035}{{\tt
  hep-ph/0304035}}].

\bibitem{Dittmaier:2003ej}
S.~Dittmaier, M.~Kramer, 1, and M.~Spira, {\it {Higgs radiation off bottom
  quarks at the Tevatron and the CERN LHC}},  {\em Phys. Rev.} {\bf D70} (2004)
  074010, [\href{http://xxx.lanl.gov/abs/hep-ph/0309204}{{\tt
  hep-ph/0309204}}].

\bibitem{Dawson:2003kb}
S.~Dawson, C.~B. Jackson, L.~Reina, and D.~Wackeroth, {\it {Exclusive Higgs
  boson production with bottom quarks at hadron colliders}},  {\em Phys. Rev.}
  {\bf D69} (2004) 074027, [\href{http://xxx.lanl.gov/abs/hep-ph/0311067}{{\tt
  hep-ph/0311067}}].

\bibitem{Maltoni:2012pa}
F.~Maltoni, G.~Ridolfi, and M.~Ubiali, {\it {b-initiated processes at the LHC:
  a reappraisal}},  {\em JHEP} {\bf 07} (2012) 022,
  [\href{http://xxx.lanl.gov/abs/1203.6393}{{\tt arXiv:1203.6393}}]. [Erratum:
  JHEP04,095(2013)].

\bibitem{Lim:2016wjo}
M.~Lim, F.~Maltoni, G.~Ridolfi, and M.~Ubiali, {\it {Anatomy of double
  heavy-quark initiated processes}},
  \href{http://xxx.lanl.gov/abs/1605.09411}{{\tt arXiv:1605.09411}}.

\bibitem{Forte:2015hba}
S.~Forte, D.~Napoletano, and M.~Ubiali, {\it {Higgs production in bottom-quark
  fusion in a matched scheme}},  {\em Phys. Lett.} {\bf B751} (2015) 331--337,
  [\href{http://xxx.lanl.gov/abs/1508.01529}{{\tt arXiv:1508.01529}}].

\bibitem{Cacciari:1998it}
M.~Cacciari, M.~Greco, and P.~Nason, {\it {The P(T) spectrum in heavy flavor
  hadroproduction}},  {\em JHEP} {\bf 05} (1998) 007,
  [\href{http://xxx.lanl.gov/abs/hep-ph/9803400}{{\tt hep-ph/9803400}}].

\bibitem{Forte:2010ta}
S.~Forte, E.~Laenen, P.~Nason, and J.~Rojo, {\it {Heavy quarks in
  deep-inelastic scattering}},  {\em Nucl. Phys.} {\bf B834} (2010) 116--162,
  [\href{http://xxx.lanl.gov/abs/1001.2312}{{\tt arXiv:1001.2312}}].

\bibitem{Buza:1996wv}
M.~Buza, Y.~Matiounine, J.~Smith, and W.~L. van Neerven, {\it {Charm
  electroproduction viewed in the variable flavor number scheme versus fixed
  order perturbation theory}},  {\em Eur. Phys. J.} {\bf C1} (1998) 301--320,
  [\href{http://xxx.lanl.gov/abs/hep-ph/9612398}{{\tt hep-ph/9612398}}].

\bibitem{Alwall:2014hca}
J.~Alwall, R.~Frederix, S.~Frixione, V.~Hirschi, F.~Maltoni, O.~Mattelaer,
  H.~S. Shao, T.~Stelzer, P.~Torrielli, and M.~Zaro, {\it {The automated
  computation of tree-level and next-to-leading order differential cross
  sections, and their matching to parton shower simulations}},  {\em JHEP} {\bf
  07} (2014) 079, [\href{http://xxx.lanl.gov/abs/1405.0301}{{\tt
  arXiv:1405.0301}}].

\bibitem{Wiesemann:2014ioa}
M.~Wiesemann, R.~Frederix, S.~Frixione, V.~Hirschi, F.~Maltoni, and
  P.~Torrielli, {\it {Higgs production in association with bottom quarks}},
  {\em JHEP} {\bf 02} (2015) 132,
  [\href{http://xxx.lanl.gov/abs/1409.5301}{{\tt arXiv:1409.5301}}].

\bibitem{Bonvini:2016fgf}
M.~Bonvini, A.~S. Papanastasiou, and F.~J. Tackmann, {\it {Matched predictions
  for the $b\bar{b}H$ cross section at the 13 TeV LHC}},
  \href{http://xxx.lanl.gov/abs/1605.01733}{{\tt arXiv:1605.01733}}.

\bibitem{Butterworth:2015oua}
J.~Butterworth et~al., {\it {PDF4LHC recommendations for LHC Run II}},  {\em J.
  Phys.} {\bf G43} (2016) 023001,
  [\href{http://xxx.lanl.gov/abs/1510.03865}{{\tt arXiv:1510.03865}}].

\bibitem{Ball:2014uwa}
{\bf NNPDF} Collaboration, R.~D. Ball et~al., {\it {Parton distributions for
  the LHC Run II}},  {\em JHEP} {\bf 04} (2015) 040,
  [\href{http://xxx.lanl.gov/abs/1410.8849}{{\tt arXiv:1410.8849}}].

\bibitem{Harland-Lang:2014zoa}
L.~A. Harland-Lang, A.~D. Martin, P.~Motylinski, and R.~S. Thorne, {\it {Parton
  distributions in the LHC era: MMHT 2014 PDFs}},  {\em Eur. Phys. J.} {\bf
  C75} (2015), no.~5 204, [\href{http://xxx.lanl.gov/abs/1412.3989}{{\tt
  arXiv:1412.3989}}].

\bibitem{Dulat:2015mca}
S.~Dulat, T.-J. Hou, J.~Gao, M.~Guzzi, J.~Huston, P.~Nadolsky, J.~Pumplin,
  C.~Schmidt, D.~Stump, and C.~P. Yuan, {\it {New parton distribution functions
  from a global analysis of quantum chromodynamics}},  {\em Phys. Rev.} {\bf
  D93} (2016), no.~3 033006, [\href{http://xxx.lanl.gov/abs/1506.07443}{{\tt
  arXiv:1506.07443}}].

\bibitem{Carrazza:2015aoa}
S.~Carrazza, S.~Forte, Z.~Kassabov, J.~I. Latorre, and J.~Rojo, {\it {An
  Unbiased Hessian Representation for Monte Carlo PDFs}},  {\em Eur. Phys. J.}
  {\bf C75} (2015), no.~8 369, [\href{http://xxx.lanl.gov/abs/1505.06736}{{\tt
  arXiv:1505.06736}}].

\bibitem{Gao:2013bia}
J.~Gao and P.~Nadolsky, {\it {A meta-analysis of parton distribution
  functions}},  {\em JHEP} {\bf 07} (2014) 035,
  [\href{http://xxx.lanl.gov/abs/1401.0013}{{\tt arXiv:1401.0013}}].

\bibitem{Watt:2012tq}
G.~Watt and R.~S. Thorne, {\it {Study of Monte Carlo approach to experimental
  uncertainty propagation with MSTW 2008 PDFs}},  {\em JHEP} {\bf 08} (2012)
  052, [\href{http://xxx.lanl.gov/abs/1205.4024}{{\tt arXiv:1205.4024}}].

\bibitem{Harlander:2011aa}
R.~Harlander, M.~Kramer, and M.~Schumacher, {\it {Bottom-quark associated
  Higgs-boson production: reconciling the four- and five-flavour scheme
  approach}},  \href{http://xxx.lanl.gov/abs/1112.3478}{{\tt arXiv:1112.3478}}.

\bibitem{Bonvini:2015pxa}
M.~Bonvini, A.~S. Papanastasiou, and F.~J. Tackmann, {\it {Resummation and
  matching of b-quark mass effects in $ b\overline{b}H $ production}},  {\em
  JHEP} {\bf 11} (2015) 196, [\href{http://xxx.lanl.gov/abs/1508.03288}{{\tt
  arXiv:1508.03288}}].

\bibitem{Anastasiou:xxx}
{\bf LHC Higgs Cross Section Working Group} Collaboration, C.~Anastasiou
  et~al., {\it {Handbook of LHC Higgs Cross Sections: 4. Deciphering the nature
  of the Higgs sector}},  \href{http://xxx.lanl.gov/abs/{\it in
  preparation}}{{\tt {\it in preparation}}}.

\end{thebibliography}\endgroup
\end{document}